\documentclass[12pt]{article}
\usepackage[applemac]{inputenc}
\usepackage{amsmath}
\usepackage{amsfonts}
\usepackage{amssymb}
\usepackage{mltex}
\usepackage{graphicx}
\newtheorem{lemma}{Lemma}[section]
\newtheorem{theorem}[lemma]{Theorem}
\newtheorem{proposition}[lemma]{Proposition}
\newtheorem{corollary}[lemma]{Corollary}
\newtheorem{remark}[lemma]{Remark}

\newtheorem{definition}[lemma]{Definition}

\def\sq{\hbox {\rlap{$\sqcap$}$\sqcup$}}
\def\sq{\hbox {\rlap{$\sqcap$}$\sqcup$}}

\def\1{{\rm 1\mskip-4.5mu l} }
\def\lsim{\raise0.3ex\hbox{$<$\kern-0.75em\raise-1.1ex\hbox{$\sim$}}}
\def\gsim{\raise0.3ex\hbox{$>$\kern-0.75em\raise-1.1ex\hbox{$\sim$}}}

\def\beq{\begin{equation}}   \def\edq{\end{equation}}
\def\bea{\begin{eqnarray}}  \def\eea{\end{eqnarray}}

\renewcommand{\theequation}{\thesection.\arabic{equation}}
\newcounter{hran} \renewcommand{\thehran}{\thesection.\arabic{hran}}

\def\bmini{\setcounter{hran}{\value{equation}}
    \refstepcounter{hran}\setcounter{equation}{0}
    \renewcommand{\theequation}{\thehran\alph{equation}}\begin{eqnarray}}

\def\bminiG#1{\setcounter{hran}{\value{equation}}
\refstepcounter{hran}\setcounter{equation}{-1}
\renewcommand{\theequation}{\thehran\alph{equation}}
\refstepcounter{equation}\label{#1}\begin{eqnarray}}

\def\emini{\end{eqnarray}\relax\setcounter{equation}{\value{hran}}\renewcommand{
\theequation}
{\thesection.\arabic{equation}}}

\begin{document}
{\title{{\bf \LARGE BLOCK-CIRCULANT MATRICES WITH CIRCULANT BLOCKS,
WEIL SUMS AND MUTUALLY
 UNBIASED BASES,\\
  II. THE PRIME POWER CASE
}}}
\author{Monique Combescure}
\maketitle

\begin{abstract}
In our previous paper \cite{co1} we have shown that the theory of circulant matrices allows
to recover the result that there exists $p+1$ Mutually Unbiased Bases in dimension $p$, $p$ being an arbitrary
prime number. Two orthonormal bases $\mathcal B,\ \mathcal B'$ of $\mathbb C^d$ are
 said mutually unbiased if 
$\forall b\in \mathcal B, \ \forall b' \in \mathcal B'$ one has that
$$\vert b\cdot b'\vert = \frac{1}{\sqrt d}$$
($b\cdot b'$ hermitian scalar product in $\mathbb C^d$). In this paper we show that the theory of block-circulant
matrices with circulant blocks allows to show very simply the known result that if $d=p^n$
($p$ a prime number, $n$ any integer) there exists $d+1$ mutually Unbiased Bases in
$\mathbb C^d$. Our result relies heavily on an idea of Klimov, Muñoz, Romero \cite{klimuro}.
As a subproduct we recover properties of quadratic Weil sums for $p\ge 3$, which generalizes the fact
that in the prime case the quadratic Gauss sums properties follow from our results.
\end{abstract}

\section{INTRODUCTION}

The Mutually Unbiased Bases in dimension $d$ are a set $\left\{\mathcal B_{1},..., \mathcal B_{N}\right\}$
of orthonormal bases in $\mathbb C^d$ such that for any $b_{j}^{(k)}\in \mathcal B_{k},
\ b_{j'}^{(k')}\in \mathcal B_{k'}$
one has
$$\left\vert b_{j}^{(k)}\cdot b_{j'}^{(k')}\right\vert= \frac{1}{\sqrt d},\ \forall j,j'=1, ..., d,\ \forall
k'\ne k= 1, ..., N$$
where $b\cdot b'= \sum_{j=1}^{d}b_{j}^*b'_{j}$ is the usual scalar product in $\mathbb C^d$.\\
This notion of mutually unbiased bases emerged in the seminal work of Schwinger \cite{sch}
and turned out to be a cornerstone in the theory of quantum information. Furthemore it is
strongly linked with the theory of Hadamard matrices \cite{ha} and to the Gauss Sums properties.\\

An important problem is the maximum number of mutually unbiased bases (MUB) in dimension $d$.
The problem has been completely solved for $d=p^n$ where $p$ is a prime number, and $n$ any
integer, in which case one can find $N=d+1$ MUB's \cite{band}\cite{tu}\cite{wo}
\cite{iv}\cite{cha}.\\

In a previous paper \cite{co1} (hereafter refered to as I) we have provided a construction of $d+1$ MUB's for $d$ a prime number using a
new method involving circulant matrices. Then the MUB problem reduces to exhibit a circulant
matrix $C$ which is a unitary Hadamard matrix, such that its powers are also circulant unitary Hadamard
matrices. Then using the Discrete Fourier Transform $F_{d}$ which diagonalizes all circulant matrices,
we have shown that a MUB in that case is just provided by the set of column vectors of the set
of matrices $\left\{ F_{d}, \1, C, C^2, ..., C^{d-1}\right\}$. Properties of quadratic Gauss sums
follow as a by-product of the method.\\

The present paper is a continuation of I, where we consider $d=p^n$. Here circulant matrices are
replaced by a set of block-circulant with circulant blocks matrices. Again the discrete Fourier transform
which in this case will be simply $F\equiv F_{p}\otimes F_{p}\otimes ...\otimes F_{p}$ will play
a central role since it diagonalizes all block-circulant with circulant blocks matrices. We follow
an idea of \cite{klimuro} to define them. The new result developed here  is that these
block-circulant matrices with circulant blocks together with $F$ will solve the MUB problem 
in that case.
\\

Let $\mathcal B_{k}= \left\{ b_{0}^{(k)}, b_{1}^{(k)}, ..., b_{d-1}^{(k)}\right\}$ 
be  orthonormal bases. Then in any given base, they are represented by  unitary matrices $B_{k}$.
  Taking
$\mathcal B_{0}$ to be the natural base, we have that
$$b_{j}^{(k)}\cdot b_{j'}^{(k')}= (B_{k}^*B_{k'})_{j,j'}$$
Thus in order that the bases $\mathcal B_{k}$ be unbiased, we just need that all the unitary 
matrices $B_{k}^*B_{k'},\ k\ne k'$ have matrix elements of modulus $d^{-1/2}$. Such
matrices are known as {\bf unitary Hadamard Matrices} (\cite{ha}).

\section{THE SQUARE OF A PRIME}

Let $p$ be a prime number. One defines a primitive $p$-th root of unity :
$$\omega= \exp\left(\frac{2\pi i}{p}\right)$$
The Discrete Fourier Transform in $\mathbb C^p$ is
$$F_{p}= \frac{1}{\sqrt p}\begin{pmatrix}1&1&1&.&.&1\\
1&\omega&\omega^2&.&.&\omega^{p-1}\\
1&\omega^2&\omega^4&.&.&\omega^{2(p-1)}\\
.&.&.&.&.&.\\
1&\omega^{p-1}&\omega^{2(p-1)}&.&.&\omega^{(p-1)(p-1)}\end{pmatrix}$$

\begin{definition}
Consider a d-periodic sequence $a_{0}, a_{1}, ..., a_{d-1}, a_{0}, a_{1},...$.\\
(i) A $d\times d$ matrix $D$ is diagonal and called ${\rm diag}(a_{0}, ..., a_{d-1}) $ if its
matrix elements satisfy
$$D_{j,k}= a_{k}\delta_{j,k},\ \forall j,k =0,1, ..., d-1$$
(ii) A $d\times d$ matrix $C$ is called circulant and denoted 
$C={\rm circ}(a_{0}, ..., a_{d-1})$ if its matrix elements satisfy
$$C_{j,k}= a_{(d-1)j+k}$$
Thus it can be written as :
$$C= \begin{pmatrix}a_{0}&a_{1}&a_{2}&.&.&a_{d-1}\\
a_{d-1}&a_{0}&a_{1}&.&.&a_{d-2}\\
.&.&.&.&.&.\\
a_{1}&a_{2}&a_{3}&.&.&a_{0}\end{pmatrix}$$
(iii) A diagonal and circulant matrix must be a multiple of the identity matrix $\1$.
(iv) A $d^2\times d^2$ matrix is said to be block-circulant if it is of the form
$$C= {\rm circ}(C_{0}, C_{1}, ..., C_{d-1})$$
where the $C_{j}$ are $d\times d$ matrices. 
\\
(v) It is block-circulant with circulant blocks if furthermore the $C_{j} $ are circulant.\\
(vi) A $d\times d$ matrix $H$ is a unitary Hadamard matrix if
$$\vert H_{j,k}\vert = d^{-1/2},\ {\rm and}\ \sum_{k=0}^{d-1}H_{j,k}^*H_{k,l}= \delta_{j,l}$$
\end{definition}

We define the following $p\times p$ unitary matrices
$$X= {\rm circ}(0,0, ...,1)$$
$$Z={\rm diag}(1,\omega, ..., \omega^{p-1})$$
They obey the $\omega$- commutation rule :

\begin{lemma}
(i) $X^p=Z^p=\1$\\
(ii)
$$ZX= \omega XZ$$
(iii) Furthemore one has
$$F_{p}XF_{p}^*=Z$$
\end{lemma}

(i) and (ii) are obvious. For a proof of (iii) see \cite{da}.

\begin{proposition}
Let $C={\rm circ}(a_{0}, ..., a_{p-1})$.\\
(i) One has
$$C=\sum_{k=0}^{p-1}a_{k}X^{p-k}$$
(ii)The discrete Fourier transform diagonalizes the circulant matrices :\\
 $$F_{p}CF_{p}^*= {\rm diag}(\tilde a_{0}, ..., \tilde a_{p-1})$$
 where
$$\tilde a_{j}= \sum_{k=0}^{p-1}a_{k}\omega^{-jk}$$
(iii)  The set of circulant $p\times p$ matrices is a commutative algebra.
\end{proposition}

Proof : 
$$F_{p}CF_{p}^*= \sum_{k=0}^{p-1}a_{k}Z^{p-k}$$
But $Z^{-k}= {\rm diag}(1, \omega^{-k}, ..., \omega^{-k(p-1)})$,
hence (ii) follows. (iii) is a consequence of (i).

\begin{corollary}
If the sequence $\left\{ a_{k}\right\}_{k\in \mathbb F_{d}}$ is such that
$$\vert a_{k}\vert = d^{-1/2}\ {\rm and}\ \vert \tilde a_{k}\vert=1$$
then $C$ is a circulant unitary Hadamard matrix.
\end{corollary}

Proof : $C$ is unitarily equivalent to an unitary matrix if $\vert \tilde a_{k}\vert=1$.
Furthemore $\vert a_{k}\vert= d^{-1/2}$, hence the result follows (see \cite{co1}).\\

\noindent
The discrete Fourier transform in $\mathbb C^{p^2}$ is defined as follows 
\beq
\label{discft}
F=F_{p}\otimes F_{p}
\edq

It has the following important property (similar to the property that the discrete
Fourier transform diagonalizes all circulant matrices) :

\begin{proposition}
(i) $F$ is an unitary Hadamard matrix.\\
(ii)
All block-circulant matrices $C$ with circulant blocks are diagonalized by $F$ :
$$FCF^*= D$$
where $D$ is a $p^2\times p^2$ diagonal matrix.
\end{proposition}

For a proof of this result see \cite{da}.\\

We shall be interested in finding block-circulant with circulant blocks unitary matrices in 
$\mathbb C^{p^2}$ that
are Hadamard matrices. An example is of course $C\otimes C'$ where $C,\ C'$  are unitary circulant Hadamard
matrices.

\medskip
\noindent
For $p$ a prime number, denote by $\mathbb F_{p}$ the field of residues modulo $p$.
The corresponding Galois field $GF(p^2)$ is defined as follows. For any $p$ there exists 
an irreducible polynomial of degree two, with coefficients in $\mathbb F_{p}$ so that if
we denote by $\alpha$ a root of this polynomial,
$$GF(p^2)= \left\{ m\alpha+n\right\}_{m,n\in \mathbb F_{p}}$$
The product $\theta\cdot \theta'\in GF(p^2)$ for $\theta,\  \theta'\in GF(p^2)$ is obtained using
the irreducible polynomial which expresses $\alpha^2$ in terms of $\alpha$ and 1.\\
The additive characters $\chi(\theta)$ in $GF(p^2)$ are defined as follows:

\begin{definition}
The additive characters on $GF(p^2)$ are :
$$\chi(\theta)= \exp\left(\frac{2i\pi}{p}{\rm tr}\theta\right)$$
where
$${\rm tr}\theta= \theta+\theta^{p}$$
\end{definition}

\begin{lemma}
They satisfy :\\
(i) $\chi(\theta+\theta')= \chi(\theta)\chi(\theta')$\\
(ii) One has :
$$\sum_{\theta\in GF(p^2)}\chi(\theta)=0 $$
(iii) 
\beq
\label{chi}
\sum_{\theta'\in GF(p^2)}\chi(\theta\cdot\theta')= p^2\delta_{\theta,0}
\edq
\end{lemma}

We take as natural basis in $\mathbb C^{p^2}$ the set of states labelled by $\theta\in GF(p^2)$, in
the following order :
$$\mathcal B\equiv \left\{\vert 0\rangle, \vert \alpha\rangle, \vert 2\alpha\rangle, ..., \vert 1 \rangle,
\vert 1+\alpha\rangle, ..., \vert p-1\rangle,..., \vert p-1+(p-1)\alpha\rangle\right\}$$

Labelled by $\theta \in GF(p^2)$ we define a set of unitary 
operators in $\mathbb C^{p^2}$ such that :

\begin{definition}
(i) The set of operators $\mathcal F \equiv\left\{X_{\theta}\right\}_{\theta\in GF(p^2)}$
obeys
\beq
\label{x}
X_{\theta}\vert \theta'\rangle
 =\vert \theta+\theta'\rangle,\ \forall \theta'\in
GF(p^2)
\edq
(ii) The set of diagonal operators $\mathcal F'\equiv\left\{ Z_{\theta}\right\}_{\theta\in GF(p^2)}$
obeys
\beq
\label{z}
Z_{\theta}\vert \theta'\rangle= \chi(\theta\cdot\theta')\vert \theta\rangle,\ 
\forall \theta'\in GF(p^2)
\edq

\end{definition}

They obey :

\begin{proposition}
(i) 
\beq
\label{comrule}
Z_{\theta}X_{\theta'}= \chi(\theta\cdot\theta')X_{\theta'}Z_{\theta}
\edq
(ii) The operators in $\mathcal F,\ \mathcal F'$ obey the group commutative property :
\beq
\label{chainrule}
X_{\theta+\theta'}= X_{\theta}X_{\theta'}=X_{\theta'}X_{\theta},\quad Z_{\theta+\theta'}= Z_{\theta}Z_{\theta'}= 
Z_{\theta'}Z_{\theta} 
\edq
(iii) $Z_{0}=X_{0}=\1$
\end{proposition}
 Proof : Take any $\varphi\in GF(p^2)$. Then
 $$Z_{\theta}X_{\theta'}\vert\varphi\rangle= \chi(\theta\cdot(\theta'+\varphi))
 \vert \theta'+\varphi\rangle= \chi(\theta\cdot\theta')X_{\theta'}Z_{\theta}
 \vert\varphi\rangle= \chi(\theta\cdot\theta')\chi(\theta\cdot\varphi)\vert\theta'
 +\varphi\rangle$$
 We also have (ii) :
 $$Z_{\theta+\theta'}\vert\varphi\rangle= \chi((\theta+\theta')\cdot\varphi)
 \vert \varphi\rangle=
 \chi(\theta\cdot\varphi)\chi(\theta'\cdot\varphi)\vert\varphi\rangle=
  Z_{\theta}Z_{\theta'}\vert \varphi\rangle$$
 \\
 \sq
 
 \begin{theorem}
 (i) $$\mathcal F= \left\{X^m\otimes X^n\right\}_{m,n\in \mathbb F_{p}}$$
 More precisely one has
 \beq
 \label{xx}
 X_{m\alpha+n}= X^n\otimes X^m
 \edq
 (ii) All operators in $\mathcal F$ are represented by unitary block-circulant with circulant blocks
 $p^2\times p^2$ matrices.\\
 (iii) 
 $$\mathcal F'= \left\{Z^m\otimes Z^n\right\}_{m,n\in \mathbb F_{p}}$$
 (iv) All operators in $\mathcal F'$ are represented by diagonal $p^2\times p^2$ matrices.\\
 (v) For any $\theta'\in GF(p^2)$ there exists a $\theta\in GF(p^2)$ such that
 \beq
 \label{xz}
 FX_{\theta'}F^*=Z_{\theta}
 \edq
 \end{theorem}
 
 Proof of (i) : It is enough to see that $X_{\alpha}= \1\otimes X$ and $X_{1}= X\otimes \1$
 since the other matrices $X_{m\alpha+n}$ will be given by the chain rule :
 $$X_{m\alpha+n}= X_{\alpha}^mX_{1}^n$$
 But these are obviously block-circulant with circulant blocks matrices. \\
 One has for $\theta'= m\alpha+n$ :
 $$FX_{m\alpha+n}F^*= FX_{\alpha}^mX_{1}^nF^*= (F_{p}\otimes F_{p})(X^n\otimes X^m)(F_{p}^*\otimes F_{p}^*)
 =Z^n\otimes Z^m$$
 which is $Z_{\theta}$ for some $\theta\in GF(p^2)$.
 \sq
 \\
 One recalls a famous result \cite{da} :
 \begin{proposition}
 All block-circulant matrices with circulant blocks commute and are diagonalized by $F$.
 \end{proposition}
Proof : It follows from (\ref{xz}) that if $C= \sum_{\theta'}\lambda_{\theta'}
X_{\theta'}$
is a block-circulant matrix with circulant blocks, one has
$$FCF^*= \sum_{\theta'\in GF(p^2)}\lambda_{\theta'}FX_{\theta'}F^*
= \sum_{\theta'\in GF(p^2)}\lambda_{\theta'}Z_{f(\theta')}$$
which is a diagonal matrix.
 
 To find the MUB's in dimension $p^2$ it is enough to exhibit a partition of the set of unitary
 operators :
 $$\mathcal E \equiv \left\{ Z_{\theta}X_{\theta'}\right\}_{\theta, \theta'\in GF(p^2)}$$
 into a set of commutant families :
 We define 
 $$\mathcal F_{0}= \mathcal F \backslash \left\{\1\right\}$$
One wants :
 $$\mathcal E= \mathcal F_{0}\bigcup_{\theta\in GF(p^2)}\mathcal C_{\theta}\cup
 \left\{\1\right\}$$
 
 The family $\mathcal C_{\theta} $ will be defined as follows :
 \begin{definition}
 Let for any $\theta \in GF(p^2)$
 $$\mathcal E_{\theta}= \left\{Z_{\theta'}X_{\theta\cdot\theta'}\right\}_{\theta'\in GF(p^2)}$$
 Define
 $$\mathcal C_{\theta}= \mathcal E_{\theta}\backslash \left\{\1\right\}$$
 \end{definition}
 
 \begin{proposition}
 (i) $\mathcal E_{0}= \mathcal F'$\\
 (ii)
 $\mathcal E_{\theta}$ is a commuting family $\forall \theta\in GF(p^2)$.
 \\
 (iii) $\mathcal E= \mathcal F_{0}\bigcup_{\theta\in GF(p^2)}\mathcal C_{\theta}\cup
 \left\{\1\right\}$ is a partition of $\mathcal E$.
 \end{proposition}
 
 Proof : (i) is obvious.\\
 (ii) 
 $\forall \theta',\ \theta''\in GF(p^2)$ one has
 $$Z_{\theta'}X_{\theta\cdot\theta'}Z_{\theta''}X_{\theta''\cdot\theta}=
 \chi(-\theta\cdot\theta'\cdot\theta'')Z_{\theta'+\theta''}X_{\theta\cdot(\theta'+\theta'')}=
 Z_{\theta''}X_{\theta\cdot\theta''}Z_{\theta'}X_{\theta\cdot\theta''}$$
 (iii) $\mathcal C_{\theta}$ and $\mathcal F_{0}$ contain $p-1$ elements. The classes
 $\mathcal C_{\theta}$ for different $\theta$ are disjoint. Therefore
 $$\bigcup_{\theta\in GF(p^2)}\mathcal C_{\theta}$$
 contains $p(p-1)$ elements.
 One has :
 $$p-1+p(p-1)+1=p^2$$
 which is the total number of elements in $\mathcal E$.
 
 \medskip
 \noindent
 Since all the unitary operators in $\mathcal C_{\theta}$ commute, they can be diagonalized by
 the same operator $R_{\theta}$. In the above cited work \cite{klimuro} they are defined
 as ``rotation operators''. In fact we shall see that they are represented in the basis $\mathcal B$
 by block-circulant with circulant block matrices. The first main result of this paper is the following :
 
 \begin{theorem}
 (i) There exists a set $\left\{R_{\theta}\right\}_{\theta\in GF(p^2)}$ of unitary 
 operators which diagonalize all the operators of the class $\mathcal C_{\theta},\ \forall 
\theta\in GF(p^2)$.\\
(ii) The operators $R_{\theta}\ {\rm for\ }\theta\ne 0$ are represented in the basis $\mathcal B$ by block-circulant with circulant block matrices
which are unitary Hadamard matrices.\\
(iii) For $p\ge 3$ they obey the group law :
$$R_{\theta+\theta'}=R_{\theta}R_{\theta'},\ \forall \theta, \ \theta'\in GF(p^2)$$
 \end{theorem} 
 
  Proof : It is enough to show that for any $\theta\in GF(p^2)\backslash \left\{0\right\}$ the $R_{\theta} $ can be expanded as
  \beq
  \label{r}
  R_{\theta}= \sum_{\theta'\in GF(p^2)}\lambda_{\theta'}^{(\theta)}X_{\theta'}
  \edq
  since they will automatically represented in the basis $\mathcal B$ by block-circulant with 
  circulant blocks matrices. One has to check that
  $$R_{\theta}^{-1}Z_{\theta'}X_{\theta\cdot\theta'}R_{\theta}=\mu_{\theta,\theta'}Z_{\theta'},\ \forall \theta'
  \in GF(p^2)$$
  Since the operators $Z_{\theta'}X_{\theta\cdot\theta'}$ are unitary, the $\mu_{\theta, \theta'}$
  are necessarily complex numbers of modulus one. But
  $$Z_{\theta'}X_{\theta\cdot\theta'}\sum_{\theta''}\lambda_{\theta''}^{(\theta)}
  X_{\theta''}= Z_{\theta'}\sum_{\theta''}\lambda_{\theta''}^{(\theta)}X_{\theta''+\theta\cdot\theta'}
  = \mu_{\theta,\theta'}\sum_{\theta'''}\lambda_{\theta'''}^{(\theta)}X_{\theta'''}Z_{\theta'}
  = \mu_{\theta, \theta'}Z_{\theta'}
  \sum_{\theta'''}\chi(-\theta'\cdot\theta''')\lambda_{\theta'''}^{(\theta)}X_{\theta'''}$$
  Equating the coefficients of $X_{\theta'''}$ in both sides we get
  $$\lambda_{\theta'''-\theta\cdot\theta'}^{(\theta)}= \mu_{\theta,\theta'}
  \chi(-\theta'\cdot\theta''')
  \lambda_{\theta'''}^{(\theta)}$$
  Taking $\theta'''=0$ and assuming that $\lambda_{0}^{(\theta)}=p^{-1},\ \forall \theta\in GF(p^2)$
  we get
  $$\lambda_{-\theta\cdot\theta'}^{(\theta)}=p^{-1} \mu_{\theta, \theta'}$$
  or equivalently, since $\theta\ne 0$ 
  $$\lambda_{\theta'}^{(\theta)}= p^{-1}\mu_{\theta, -\theta^{-1}\cdot\theta'}$$
  This proves that all the $\lambda_{\theta'}^{(\theta)}$ must be of modulus $p^{-1}$.\\
  Therefore since all the $X_{\theta}$ are represented by unitary matrices that have non-zero elements 
  (actually $1$) where all the others have zeros, and since every matrix element of $R_{\theta}$
  is of the form $\lambda_{\theta'}^{(\theta)}$ for some $\theta'\in GF(p^2)$, 
  this proves that all the $R_{\theta}$ are represented by  Hadamard matrices.\\
  Now we have to check the compatibility condition. We reexpress it in terms of
   $\mu_{\theta, \theta'}$.
  Supressing the index $\theta$ in the $\mu_{\theta, \theta'}$ for simplicity, we need
  to have $\forall \theta',\ \theta''\in GF(p^2)$ 
  $$\mu_{-\theta^{-1}(\theta''-\theta\theta')}= \mu_{\theta'}\mu_{-\theta^{-1}\theta''}
  \chi(-\theta'\cdot\theta'')$$
  or in other terms 
  \beq
  \label{mu}
  \mu_{\theta'+\theta''}= \mu_{\theta'}\mu_{\theta''}
  \chi(\theta\cdot\theta'\cdot\theta'')
  \edq
  But this results easily from the group property of the $X_{\theta}$'s and $Z_{\theta}$'s 
  (\ref{x}, \ref{z}) :
  $$R_{\theta}^{-1}Z_{\theta'+\theta''}X_{\theta\cdot(\theta'+\theta'')}R_{\theta}
=\chi(\theta\theta'\theta'')R_{\theta}^{-1}Z_{\theta'}X_{\theta\theta'}R_{\theta}
R_{\theta}^{-1} Z_{\theta''}X_{\theta\theta''}R_{\theta}= \chi(\theta\theta'\theta'')
\mu_{\theta, \theta'}\mu_{\theta, \theta''}Z_{\theta'+\theta''}$$ 
In \cite{klimuro} it is shown that for $p\ge 3$ the solution of (\ref{mu}) with $\mu_{\theta,0}=1$
is
\beq
\label{mu1}
\mu_{\theta, \theta'}= \chi(2^{-1}\theta\cdot\theta'^2)
\edq
Thus we deduce that
\beq
\label{lambda}
\lambda_{\theta'}^{(\theta)}=p^{-1} \chi(2^{-1}\theta^{-1}\cdot (\theta')^2)
\edq
We now prove the unitarity of $R_{\theta}$. For $\theta=0$ this is obvious since
$R_{0}= \1$. It is enough to check that for $\theta\ne 0$ one has :
$$\sum_{\theta'\in GF(p^2)}(\lambda_{\theta'}^{(\theta)})^*\lambda_{\theta'+\theta''}^{(\theta)}
= \delta_{\theta'', 0}$$
One has :
$$\sum_{\theta'\in GF(p^2)}(\lambda_{\theta'}^{(\theta)})^*\lambda_{\theta'+\theta''}^{(\theta)}
=\frac{1}{p^2}\sum_{\theta'\in GF(p^2)}\mu_{\theta, -\theta^{-1}\cdot \theta'}^*
\mu_{\theta, -\theta^{-1}\cdot(\theta'+\theta'')}$$
$$= \frac{1}{p^2}\sum_{\theta'\in GF(p^2)}\vert \mu_{\theta, -\theta^{-1}\cdot\theta'}\vert^2
\mu_{\theta, -\theta^{-1}\cdot \theta''}\chi(\theta^{-1}\theta'\theta'')= 
\mu_{\theta,-\theta^{-1}\cdot \theta''}
\frac{1}{p^2}\sum_{\theta'\in GF(p^2)}\chi(\theta^{-1}\theta'\theta'')
=\delta_{\theta'',0}$$
where we have used (\ref{mu}) and (\ref{chi}), together with the fact that
$\theta^{-1}\cdot\theta''=0$ implies $\theta''=0\ $
due to the field property of $GF(p^2)$. Thus one has
$$R_{\theta}^{-1}= R_{\theta}^*,\ \forall \theta\in GF(p^2)$$
\\

\noindent
(iii) The group law $R_{\theta+\theta'}=R_{\theta}R_{\theta'}$ for $p\ge 3$ 
 has been established in \cite{klimuro}.  For $p=2$ it has to be suitably modified
as shown in \cite{klimuro} (see remark below). Let us see how it works for $p\ge 3$ :
$$(R_{1}^*)^n(R_{\alpha}^*)^mZ_{\theta'}X_{(n+m\alpha)\cdot \theta'}R_{1}^n
R_{\alpha}^m
= (R_{1}^* )^n(R_{\alpha}^*)^mZ_{\theta'}X_{\theta'\cdot m \alpha}R_{\alpha}^m
X_{n\theta'}
R_{1}^n= \mu(m\alpha, \theta')(R_{1}^*)^nZ_{\theta'}X_{n\theta'}R_{1}^n
$$
$$= \mu_{m\alpha, \theta'}
\mu_{n,\theta'}Z_{\theta'}
= \mu_{m\alpha+n, \theta'}Z_{\theta'}$$
holds for $p\ge 3$ since
$$\mu_{m\alpha+n, \theta'}= \mu_{m\alpha, \theta'}\mu_{n,\theta'}, \ \forall \theta'\in GF(p^2)$$
which easily follows from (\ref{mu1}) for $p\ge 3$, and the additivity of the characters.

\begin{remark}
For $p=2$ the group law is not satisfied (see \cite{klimuro}). One has instead a very similar
property (modified group law ) :
$$R_{\alpha}R_{\alpha+1}= R_{\alpha+1}R_{\alpha}= R_{1}$$
$$R_{\alpha}R_{1}= R_{\alpha+1}X_{\alpha+1}$$
$$R_{\alpha+1}R_{1}= R_{\alpha}X_{\alpha}$$
$$R_{\alpha}^2= X_{\alpha+1}$$
$$R_{\alpha+1}^2=X_{\alpha}$$
$$R_{1}^2= X_{1}$$ 
\end{remark}
 The second main result of this paper is the following :
 
 \begin{theorem}
 The set of operators $\left\{  F, R_{\theta}\right\}_{\theta\in GF(p^2)}$
  defines a set of $p^2+1$ MUB's in $\mathbb C^{p^2}$.
 
 \end{theorem}
 
 Proof : Each $R_{\theta}$ is represented by an unitary Hadamard matrix, and so is $F$. Due to
 the group property, $R_{\theta}^*R_{\theta'}= R_{\theta'-\theta}$ thus is an unitary
 Hadamard matrix. It remains to show that $R_{\theta}F^*$ is an unitary Hadamard matrix
 $\forall \theta\in GF(p^2)$. But we have
 $$R_{\theta}F^*= F^*D_{\theta},\quad \forall \theta\in GF(p^2)$$
 with $D_{\theta}$ an unitary diagonal matrix, since $F$ diagonalizes all block-circulant
 matrices with circulant blocks. The product of $D_{\theta}$ with the
 unitary Hadamard matrix $F^*$ is obviously an unitary Hadamard matrix.\\
 In the case of dimension $2^2$ one has instead of the group property that
 $$\forall \theta,\  \theta',\quad \exists \theta''\ {\rm such\ that\ }R_{\theta}^*
 R_{\theta'}= R_{\theta'-\theta}X_{\theta''}$$
 Since $X_{\theta''}$ has exactly one non-vanishing element ( $1$) on each line and column,
 the product $R_{\theta'-\theta}X_{\theta''}$ is indeed an unitary Hadamard matrix.
 \sq
 \\
 \section{THE CASE $d=p^n$}
 
 The case $d=p^n$ with general power $n$ can be treated similarly. There is a notion of block-circulant matrices
 with block-circulant blocks and building block $p\times p$ matrices which are circulant
 which generalizes the case $n=2$. These are diagonalized by the Discrete Fourier Transform $F$ in 
 $\mathbb C^{p^n}$ which is
 $$F= F_{p}\otimes F_{p}\otimes ... \otimes F_{p}$$
 ($n$  times) which is obviously an unitary Hadamard matrix. \\
 The Galois field $GF(p^n)$ is defined through the irreducible polynomial which is of power $n$, with coefficients
 in $\mathbb F_{p}$. Thus elements of the Galois field $GF(p^n)$ are of the form
 $$\theta =\sum_{i=0}^{n-1}c_{i}\alpha^{i}$$
 where $c_{i}\in \mathbb F_{p}$ and $\alpha$ is a root of the characteristic polynomial.\\
 The characters are
 $$\chi(\theta)= \exp\left(\frac{2i\pi}{p}{\rm tr}(\theta)\right)$$
 where
 $${\rm tr}(\theta)= \theta+\theta^p+ ... + \theta^{p^{n-1}}$$
  For any $\theta\in GF(p^n)$
 the operators $X_{\theta},\ 
 Z_{\theta},\ R_{\theta}$ are defined in \cite{klimuro} and the $R_{\theta}$ obey 
 a group law (resp. a modified group law) if $p\ge 3$ (resp. $p=2$).
 \\
 As previously we have $X_{\theta}\in \left\{ X^{k_{1}}\otimes X^{k_{2}}\otimes ... \otimes
 X^{k_{n}},\ {\rm with\ } k_{j}\in \mathbb F_{p}\right\}$ and
 $$Z_{\theta}\in \left\{ Z^{k_{1}}\otimes Z^{k_{2}}\otimes ... \otimes Z^{k_{n}}\right\}_{k_{i}\in \mathbb F_{p}}$$
 and
 $$R_{\theta}= \sum_{\theta'\in GF(p^n)}\lambda_{\theta'}^{(\theta)}X_{\theta'}$$
 with $\lambda_{\theta'}^{(\theta)}$ of modulus $\frac{1}{\sqrt {p^n}}$. The
 $\lambda_{\theta'}^{(\theta)}$ are given for $p\ge 3$ similarly to (\ref{lambda}) by
 \beq
 \label{lambda'}
 \lambda_{\theta'}^{(\theta)}= p^{-n/2}\chi(2^{-1}\theta^{-1}(\theta')^2)
 \edq
  All the results of the
 previous section are easily generalized.
 
 \begin{theorem}
 The unitary Hadamard matrices  $F, R_{\theta}\ {\rm for \ }\theta\in GF(p^n)$ define a set of $p^n+1$
 MUB's in $\mathbb C^{p^n}$.
 \end{theorem}
 
 \section{Weil sums for $d=p^n, \ p\ge 3$}
 
 The Weil sums in dimension $p^n$ are the equivalent of the Gauss sums for $d=p$ ($p$ prime
 number). The characters $\chi(\theta), \ \theta\in GF(p^n)$ replace the powers 
 $\omega^n,\ n\in \mathbb F_{p}$. Usually in the literature (see for example \cite{wo}), the Weil sums properties are
 used to solve the MUB problem. Here, as in \cite{co1}, we do the converse. In the previous
 sections we have constructed the $d+1$ bases, and we shall {\bf deduce the Weil sums properties}
 from this construction.
 
 \begin{theorem}
 Let $p\ge 3$. Then for any $\theta \in GF(p^n)\backslash \left\{0\right\}$ and
 any $\theta'\in GF(p^n)$ we have
 \beq
 \label{weil}
 \left\vert\sum_{\theta''\in GF(p^n)}\chi(\theta\cdot (\theta'')^2+ \theta'\cdot\theta'')\right\vert
 = \sqrt{p^n}
 \edq
 \end{theorem} 
 
 Proof : The matrix elements of a given row in $F$ are of the form
 $$\frac{1}{\sqrt{p^n}}\chi(\theta'\cdot\theta'')_{\theta''\in GF(p^n)}$$
 for some $\theta'\in GF(p^n)$. Since 
 $$R_{\theta}= \frac{1}{\sqrt{p^n}}\sum_{\theta''\in GF(p^n)}\chi((2\theta)^{-1}\cdot (\theta'')^2)
 X_{\theta''}$$
 the matrix elements of the first column of  the matrices
 $R_{\theta},\ \theta\ne 0$ are of the form 
 $$\frac{1}{\sqrt{p^n}}\chi(2^{-1}\theta^{-1}\cdot(\theta'')^2)_{\theta''\in GF(p^n)}$$
 where we have used (\ref{lambda'}).
 But the elements $(2\theta)^{-1}$ when $\theta\in GF(p^n)\backslash \left\{0\right\}$
 span all $\theta_{1}\in GF(p^n)\backslash\left\{0\right\}$. We have established that the matrix
 $FR_{\theta}$ is Hadamard. This implies that all its matrix elements have modulus
 $p^{-n/2}$. All matrix elements of the first column of $FR_{\theta}$ are thus of the form
 $$p^{-n}\sum_{\theta''\in GF(p^n)}\chi(\theta_{1}\cdot(\theta'')^2+ \theta'\cdot \theta'')$$
 with $\theta_{1}= (2\theta)^{-1}$. Writing that its modulus is $p^{-n/2}$ yields
 equ. (\ref{weil}).

\end{document}